# Probabilistic Databases with MarkoViews[*]

Abhay Jha
Computer Science and Engineering
University of Washington
Seattle, WA 98195–2350
abhaykj@cs.washington.edu

Dan Suciu
Computer Science and Engineering
University of Washington
Seattle, WA 98195–2350
suciu@cs.washington.edu

## ABSTRACT

Most of the work on query evaluation in probabilistic databases has focused on the simple tuple-independent data model, where tuples are independent random events. Several efficient query evaluation techniques exists in this setting, such as safe plans, algorithms based on OBDDs, tree-decomposition and a variety of approximation algorithms. However, complex data analytics tasks often require complex correlations, and query evaluation then is significantly more expensive, or more restrictive.

In this paper, we propose MVDB as a framework both for representing complex correlations and for efficient query evaluation. An MVDB specifies correlations by views, called MarkoViews, on the probabilistic relations and declaring the weights of the view's outputs. An MVDB is a (very large) Markov Logic Network. We make two sets of contributions. First, we show that query evaluation on an MVDB is equivalent to evaluating a Union of Conjunctive Query(UCQ) over a tuple-independent database. The translation is exact (thus allowing the techniques developed for tuple independent databases to be carried over to MVDB), yet it is novel and quite non-obvious (some resulting probabilities may be negative!).

This translation in itself though may not lead to much gain since the translated query gets complicated as we try to capture more correlations. Our second contribution is to propose a new query evaluation strategy that exploits offline compilation to speed up online query evaluation. Here we utilize and extend our prior work on compilation of UCQ. We validate experimentally our techniques on a large probabilistic database with MarkoViews inferred from the DBLP data.

## 1. INTRODUCTION

The task of analyzing and extracting knowledge from large datasets often requires probabilistic inference over a complex probabilistic model on the data. This step represents a major challenge. Most of the scalable query processing techniques developed for probabilistic databases assume that the tuples are independent events, or disjoint-independent [4, 1, 24]. For example, MystiQ, MayBMS, and SPROUT report running times of a few seconds on databases of tens of millions of tuples, using a combination of techniques such as *safe plans* [7], plan re-orderings and functional dependencies [23], Monte-Carlo simulation combined with top-$k$ optimization [28], or approximate confidence computation that tradeoff precision for performance [24]. These systems scale to quite large databases, but are limited to independent probabilistic databases, or disjoint-independent.

Tuple-independent probabilistic databases are insufficient for analyzing and extracting knowledge from practical datasets. As has been shown in the Machine Learning community, modeling correlations is critical in complex knowledge extraction tasks. For example, in Markov Logic Networks (MLN) [29], users can assert arbitrary probabilistic statements over the data, in the form of First Order Logic sentences, and assign a weight. The sentence, called a *feature*, is expected to hold to a degree indicated by the weight. Each feature may introduce correlations between a large number of base facts, and thus the MLN can express, very concisely, a large Markov Network. MLNs have been demonstrated to be effective at a variety of tasks, such as Information Extraction [26], Record Linkage [31], Natural Language Processing [27]. A benefit of MLNs is that the same framework can be used both for learning the weights, and for inferring probabilities of new queries.

In this paper we present a new approach for representing and querying probabilistic databases. Our data model combines probabilistic databases with MLNs: it consists of a collection of probabilistic (tuples are annotated with a probability) and deterministic tables, and a collection of views, called MarkoViews. A MarkoView is expressed by a Union of Conjunctive Queries (UCQ) over the probabilistic and deterministic tables, and associates a weight to each tuple in the answer; intuitively, it asserts a likelihood for that output tuple, and therefore introduces a correlation between all contributing input tuples. A MarkoView can be seen as a set of MLN features, and thus, its weights can be learned as in MLNs; we do not address learning in this paper, but focus solely on inference, or query evaluation. We call a database consisting of probabilistic tables and MarkoViews an MVDB. The data model of MVDBs is significantly richer than that of tuple-independent probabilistic databases, which we denote with INDB.

We make two sets of contributions. First, we show how

[*]This work was supported in part by the NSF grants IIS-1115188, IIS-0915054, IIS-0911036

to translate query evaluation over an MVDB into query evaluation over an INDB. More precisely, we express the probability $P(Q)$ on an MVDB in terms of the probability $P_0(Q \vee W)$, where $W$ is a union of queries, one for each MarkoView; therefore $W$ is a UCQ. To be precise, we show that $P(Q) = P_0(Q \mid \neg W) = (P_0(Q \vee W) - P_0(W))/(1 - P_0(W))$. The probability $P_0$ is on an INDB obtained from the MVDB through a simple, yet quite non-obvious transformation, discussed in Sect. 3. Without going into the transformation details, we would like to mention that $\neg W$ logically stands for the MarkoViews, hence intuitively the translation is computing the probability of the query given that the views hold. Note that if $Q$ is a UCQ, so is the translated query; hence, both the query and the probabilistic model are simple and come from well-understood domains. In particular, while there are very few results on tractability of MLNs and none complete, the set of tractable UCQ over INDB is already known [8]. Therefore, the translation moves our problem into a domain that is well-understood and allows for easier detection of tractable instances. We are aware of one more such translation [12] in the literature, but it leads to a more complicated query, which is no longer a UCQ. In contrast, our translation leads to both a simple query (UCQ) and a simple model (tuple independent), where the complexity is well understood and the tractable cases are fully characterized.

Our second set of contributions is to devise an efficient query evaluation method for MVDB. The task is to compute $P_0(Q \vee W)$, where both $Q$ and $W$ are UCQs. Note that $W$ depends only on the MarkoViews, and does not depend on the query $Q$. We describe a new index structure for $W$, called an MV-index, and show how to use it to compute $P_0(Q \vee W)$. The MV-index consists of an Ordered Binary Decision Diagram (OBDD) [5], extended with additional information that is critical for computing $P_0(Q \vee W)$ efficiently. In prior work [15] we have shown that a certain class of UCQ queries, called *inversion-free* queries, always have an OBDD whose size is linear in the size of the active domain of the database. Here, we extend that construction to an algorithm that constructs an OBDD for *any* $W$: in the particular case when $W$ is inversion-free, the resulting OBDD is guaranteed to be linear in the size of the database. The OBDD itself is not sufficient for computing $P_0(Q \vee W)$ efficiently: state of the art synthesis algorithms require a time proportional to the product of the sizes of the two OBDDs for $Q$ and $W$ respectively. We describe here how to extend the OBDD to an MV-index, then present a top-down evaluation algorithm for computing $P_0(Q \vee W)$, which traverses only a small fraction of the MV-index.

**Running Example** We illustrate in Fig. 1 how MarkoViews can be used to add domain knowledge to DBLP [19], a database consisting of a few million tuples. We use an approach developed in the AI community for inferring new relations, such as *advisor*, or *affiliation*, from a database of citations [29, 30, 18]. Any MVDB has three parts.

First, the deterministic database. In our example its is described at the top of Fig. 1, and consists of four base tables, Author, Wrote, Pub, HomePage, and two materialized views (FirstPub(aid,year), which records for each author the year of her first publication, and DBLPAffiliation(aid,inst), which associates an affiliation to some authors[1]).

Second, an MVDB has probabilistic tables, shown in the middle of Fig. 1: $\text{Student}^p(\text{aid,year})$ stores likely years when an author was a student, $\text{Advisor}^p(\text{aid1,aid2})$ stores likely advisor/advisee relationship, and $\text{Affiliation}^p(\text{aid,inst})$ records inferred affiliations based on co-authorship. Each probabilistic table is defined by a query, which also associates a weight to every output tuple; for example, $\text{Student}^p(\text{aid, year})[e^{1-.15(\text{year}-\text{year'})}] : - \ldots$ associates the weight $w = e^{1-.15(\text{year}-\text{year'})}$ to its output. Weights are often preferred over probabilities when the probability function is a product of potential functions, as in MLNs and MVDBs. The intuition is that the weight $w$ represents the odds of a probability $p$, $w = p/(1-p)$ (formal definition in Sect. 2).

Third, the MVDB contains a set of MarkoViews, which in our example are shown at the bottom of Fig. 1. Each MarkoView is a query over probabilistic tables, and its purpose is to define some correlations between the tuples in those tables. It does this by defining a view over the probabilistic tables, then asserting a certain weight for the tuples in the view. Weights $< 1$ define a negative correlation, weights $> 1$ define a positive correlation, and a weight $= 1$ means independence. A weight $= 0$ means a hard constraint: the view must be empty. For example, the MarkoView V1 defines a correlation between a tuple in $\text{Student}^p$ and a tuple in $\text{Advisor}^p$: it states that the more papers two people co-author during the years when the second person was a student, the more likely that the first person was his advisor. V2 defines a hard constraint: each person can have only one advisor. Finally, V3 introduces positive correlations between common affiliations for people who published a lot together.

Consider now the following simple query on the MVDB: *find all students advised by Sam Madden*. The query, written over the MVDB, is shown in Fig. 2 (a). If the tuples in $\text{Student}^p$ and $\text{Advisor}^p$ were independent random variables, then this query could be computed very efficiently, because it is a safe query [7]. However, MarkoViews introduce correlations between the probabilistic tuples. We show in this paper that the probability of an answer aid, $P(\text{Q}(\text{aid}))$, can be expressed in terms of the probability of a query $P_0(\text{Q}(\text{aid}) \vee \text{W})$ over a tuple independent database. We give the exact formula in Fig. 2 (c). The new INDB has five tuple-independent probabilistic tables: $\text{Student}^p$ and $\text{Advisor}^p$ (which have the same sets of possible tuples as in the MVDB, and with the same weights), and $\text{NV1}^p$, $\text{NV2}^p$, $\text{NV3}^p$, which are three new, tuple-independent probabilistic tables, whose possible tuples are obtained from the MarkoViews V1, V2, V3, and whose weights are derived from the latter through the formula $(1-w)/w$. Note that if $w > 1$, then the translated weight is negative, which, in turn, corresponds to a negative probability; thus, the INDB may have some tuples with negative probabilities! However, the expression for $P(\text{Q})$ is exact, hence its value is guaranteed in $[0, 1]$. We discuss the translation from MarkoViews to INDBs in Sect. 3, and, in particular, the implications of having negative probabilities in a database in Sect. 3.3.

Query evaluation on an MVDB reduces to evaluating formula like Fig. 2 (c), on an INDB. The query dependent part of this expression is $P_0(\text{Q}(\text{aid}) \vee \text{W})$. While this requires standard evaluation of a query over a tuple-independent data-

---

[1]We computed the institute from the person's Webpage, when it was available in DBLP. For example, both Luis Gravano www.cs.columbia.edu/~gravano and Ken Ross www.cs.columbia.edu/~kar have the same institute, www.cs.columbia.edu.

| Tables obtained from DBLP | |
|---|---|
| Table | # Tuples |
| `Author(aid, name)` | 1M |
| `Wrote(aid, pid)` | 4.5M |
| `Pub(pid, title, year)` | 1.7M |
| `HomePage(aid, url)` | 18.7K |

| Derived tables (standard views) | |
|---|---|
| Table | # Tuples |
| `FirstPub(aid,year)` | 1M |
| `DBLPAffiliation(aid,inst)` | 18.7K |

Three probabilistic tables $\text{Student}^p$, $\text{Advisor}^p$, $\text{Affiliation}^p$.

| Possible Tuples | Description | Size |
|---|---|---|
| `Student`$^p$`(aid,year)[exp(1-.15*(year-year'))] :-`<br>`    FirstPub(aid,year'), year' - 1 <= year <= year' + 5` | `aid` was a student in that year if his first publication was not long before. | 6M |
| `Advisor`$^p$`(aid1,aid2)[exp(.25*count(pid))] :-`<br>`    Student(aid1,year), Wrote(aid1,pid), Wrote(aid2,pid),`<br>`    Pub(pid,title,year), not Student(aid2,year),`<br>`    count(pid) > 2` | `aid2` was `aid1`'s advisor if they published enough papers together while `aid1` was a student and `aid2` was not. | .25M |
| `Affiliation`$^p$`(aid,inst)[exp(.1*count(pid))] :-`<br>`    Wrote(aid,pid), Wrote(aid2,pid),`<br>`    DBLPAffiliation(aid2,inst), Pub(pid,title,year),`<br>`    aid<>aid2, year>2005, not DBLPAffiliation(aid,inst2)` | `aid`'s affiliation is `inst` if she published recently with people from `inst` | .27M |

The MARKOVIEWS in the MVDB

| View definition | Description | Size |
|---|---|---|
| `V1(aid1,aid2)[count(pid)/2] :- Advisor`$^p$`(aid1,aid2),`<br>`    Student`$^p$`(aid1,year), Wrote(aid1,pid),`<br>`    Wrote(aid2,pid), Pub(pid,title,year)` | The more they published together while `aid2` was a student, the more likely `aid1` was his advisor | .25M |
| `V2(aid1,aid2,aid3)[0] :- Advisor`$^p$`(aid1,aid2),`<br>`    Advisor`$^p$`(aid1,aid3), aid2 <> aid3` | A person has only one advisor | .38M |
| `V3(aid1,aid2,inst)[count(pid)/5] :-`<br>`    Affiliation`$^p$`(aid1,inst), Affiliation`$^p$`(aid2,inst),`<br>`    Wrote(aid1,pid), Wrote(aid2,pid), Pub(pid,title,year),`<br>`    year > 2004, count(pid) > 30` | If two people have published a lot together recently, then their affiliations are very likely to be same | 1.5K |

**Figure 1: An illustration of MarkoViews on the DBLP database**

base, it can still be a major challenge: the reason is that the lineage of W is usually large, because it includes most probabilistic tuples and all tuples in the MARKOVIEWS. By contrast, the lineage of Q is much smaller, because it has a selection predicate ("`%Madden%`"). This is why we use the strategy of compiling W offline. More precisely, we construct an MV-index, by first converting W into an OBDD, then adding appropriate pointer structures. Using the MV-index, query evaluation can be sped up dramatically: Q took 15 ms and returned results for all 48 advisors similarly named. We describe query compilation in Sect. 4, and describe experiments in Sect. 5.

## 2. DEFINITIONS

We use $\mathbf{R}$ to denote a relational schema with relation names $R_1, R_2, \ldots, R_k$. We assume each relation has a key.[2] A database instance $I$ is a $k$-tuple $(R_1^I, \ldots, R_k^I)$, where $R_i^I$ is an instance of the relation $R_i$; with some abuse of notation we drop the superscript $I$ and write $R_i$ for both the relation name and the instance of that relation.

### 2.1 Probabilistic Databases

A probabilistic database is a pair $\mathbf{D} = (\mathbf{W}, P)$, where $\mathbf{W} = \{I_1, \ldots, I_N\}$ is a set of instances, called *possible worlds*, and $P : \mathbf{W} \to [0, 1]$ is a function such that $\sum_{j=1,N} P(I_j) = 1$. Thus, the instance is not known with certainty: every possible world $I_j$ has some probability, $P(I_j)$. A relation $R_i$ is called *deterministic* if it has the same instance in all possible worlds $R_i^{I_1} = \cdots = R_i^{I_N}$; otherwise, we say that $R_i$ is probabilistic, and we sometimes add the superscript $p$, writing $R_i^p$ to indicate that $R_i$ is probabilistic. For example in Fig. 1, the deterministic relations are `Author`, `Wrote`, `Pub`, `HomePage`, and the probabilistic relations are $\text{Student}^p$, $\text{Advisor}^p$, $\text{Affiliation}^p$.

---
[2]As usual, if there is no natural key, then the set of all attributes constitutes a key.

We denote **Tup** the set of *possible tuples*, i.e. the set of all tuples occurring in all possible worlds $I_1, \ldots, I_N$. The tuples in **Tup** include the relation name where they come from, e.g. the tuples $R(a,b)$ and $S(a,b)$ are considered distinct tuples in **Tup**. We associate to each tuple $t \in$ **Tup** a Boolean random variable, denoted $X_t$: given a random world $I_j$, $X_t = 0$ if $t \notin I_j$ and $X_t = 1$ if $t \in I_j$. The probability of the event $\exists j. \ t \in I_j$ is denoted $P(t)$ or $P(X_t)$, and is also called the marginal probability of the tuple $t$.

A query is denoted $Q(\bar{x})$, where $\bar{x}$ are called free variables, or head variables. The answer to $Q$ on an instance $I$, $Q(I)$, is the set of all tuples $\bar{a}$ s.t. $I \models Q(\bar{a})$, where $Q(\bar{a})$ is the Boolean query obtained by substituting the head variables $\bar{x}$ with the constants $\bar{a}$. The answer on a probabilistic database, $\mathbf{D}$, is a set of pairs of the form $(\bar{a}, p)$, where $\bar{a} \in Q(I)$ for some possible world $I$, and:

$$p = P(Q(\bar{a})) = \sum_{i:\bar{a} \in Q(I_i)} P(I_i)$$

```
Q(aid) :- Student^p(aid), Advisor^p(aid,aid1),
         Author^p(aid,n), Author^p(aid1,n1),
         n1 like '%Madden%'
```
(a)

```
W1 :- NV1^p(aid1,aid2), Advisor^p(aid1,aid2),
      Student^p(aid1,year), Wrote(aid1,pid),
      Wrote(aid2,pid), Pub(pid,title,year)
W2 :- NV2^p(aid1,aid2,aid3), Advisor^p(aid1,aid2),
      Advisor^p(aid1,aid3), aid2 <> aid3
W3 :- NV3^p(aid1,adi2,inst), Affiliation^p(aid1,inst),
      Affiliation^p(aid2,inst), Wrote(aid1,pid),
      Wrote(aid2,pid), Pub(pid,title,year),
      year > 2004, count(pid) > 30
W  :- W1 ∨ W2 ∨ W3
```
(b)

$$P(\mathtt{Q(aid)}) = \frac{P_0(\mathtt{Q(aid)} \vee \mathtt{W}) - P_0(\mathtt{W})}{1 - P_0(\mathtt{W})}$$
(c)

**Figure 2: A query Q over the MVDB (a); helper queries on the INDB(b); expressing the probability on the MVDB in terms of the probability on the INDB (c).**

The queries we consider in this paper are *Unions of Conjunctive Queries*, denoted UCQ, which are expressions of the form $Q(\bar{x}) = Q_1(\bar{x}) \vee \ldots \vee Q_m(\bar{x})$, where each $Q_i(\bar{x})$ is a conjunctive query, i.e. has the form $\exists \bar{y}_i . \varphi_i(\bar{x}, \bar{y}_i)$ where $\varphi_i$ is a conjunction of positive, relational atoms, and/or inequality predicates, such as $z < 5$. We write queries in datalog notation, indicating the head variables. For example, $Q(x) = R(x), S(x,y)$ denotes the query $\exists y. R(x) \wedge S(x,y)$, while $Q = R(x), S(x,y)$ denotes the Boolean query $\exists x. \exists y. R(x) \wedge S(x,y)$. With some abuse of notation, we allow the use of aggregates and negations, but only on the deterministic tables[3].

### 2.2 Tuple-Independent Databases

A probabilistic database is *tuple-independent* if, for any set of possible tuples $t_1, t_2, \ldots, t_n$, the events $X_{t_1}, X_{t_2}, \ldots, X_{t_n}$ are independent. We write $\mathbf{D}_0$ for a tuple-independent database, and also denote it with INDB. It is uniquely defined by a pair $(\mathbf{Tup}, p)$, where $\mathbf{Tup}$ is the set of possible tuples and $p : \mathbf{Tup} \to [0, 1]$ is any function. The possible worlds are all subsets $I \subseteq \mathbf{Tup}$, and their probabilities are $P(I) = \prod_{t \in I} p(t) \cdot \prod_{t \in \mathbf{Tup} - I} (1 - p(t))$.

Tuple-independent probabilistic databases are the simplest, and most intensively studied types of probabilistic databases [33]. Even though the input tuples are independent, correlations are introduced during query evaluation, and query evaluation is, in general, #P-complete, e.g. for the Boolean query $Q = R(x), S(x, y), T(y)$. However, these probabilistic databases are now well understood, and a complete characterization of UCQ queries into #P-complete and PTIME queries exists [8]. In addition, many practical methods for query evaluation on tuple-independent databases have been proposed in the literature (see Sect. 6).

---

[3]For example we used count(pid) in V3 in Fig. 1, meaning that the subquery consisting of the last two lines is first computed as a view over the deterministic tables, then the resulting view is used as a single table in V3; after this transformation, V3 becomes a conjunctive query over probabilistic and deterministic tables.

### 2.3 Markov Logic Networks (MLNs)

A Markov Logic Network is a set $L = \{(F_1, w_1), \ldots, (F_m, w_m)\}$, where each $F_i$ is a formula in First Order Logic called a *feature*, and each $w_i$ is a number called the *weight* [29, 9]. The formulas are over a relational vocabulary $\mathbf{R}$, and may have free variables, which are interpreted as being universally quantified. Let $C$ be a finite set of constants. A *grounding* of a formula $F_i$ is a formula of the form $F_i[\bar{a}/\bar{x}]$, where the free variables $\bar{x}$ of $F_i$ are substituted with some constants $\bar{a}$ in $C$; let $G(F_i)$ denote the set of grounding of $F_i$, and let $G(L) = \{(G, w_i) \mid \exists (F_i, w_i) \in L : G \in G(F_i)\}$ be the *grounded MLN*. Let $\mathbf{Tup}$ be the set of ground tuples with the relation symbols in $\mathbf{R}$ and constants in $C$. The *weight* $\Phi(I)$ of a world $I \subseteq \mathbf{Tup}$, and the *partition function* $Z$ are:

$$\Phi(I) = \prod_{(G,w) \in G(L): I \models G} w \quad (1)$$

$$Z = \sum_{I \subseteq \mathbf{Tup}} \Phi(I) \quad (2)$$

DEFINITION 1. *The semantics of an MLN $L$ is the probabilistic database $\mathbf{D}_L = (\mathbf{W}, P)$, where $\mathbf{W} = \{I \mid I \subseteq \mathbf{Tup}\}$ and $P(I) = \Phi(I)/Z$ for all $I \subseteq \mathbf{Tup}$.*

The intuition is the following. Any subset of tuples is a possible world, and its weight is the product of the weights of all grounded features that are true in that world. The probability is obtained by normalizing with $Z$. Note that a feature weight $w > 1$ means that worlds where the feature holds are more likely; $w < 1$ means that worlds were the feature holds are less likely; and $w = 1$ means indifference. A weight $w = \infty$ is interpreted as a hard constraint: only worlds that satisfy the feature are considered as possible. This can be seen by letting $w \to \infty$ in the expression $P(I) = \Phi(I)/Z$.

MLNs have been used in several applications of Machine Learning [31, 26, 27, 9]. One reason for their popularity is that they use the same formalism for both learning (of the weights $w$) and for probabilistic inference. There are two types of inferences within MLNs: MAP (maximum a posteriori) inference, which computes the most likely world, and marginal inference, which sums the probabilities of all worlds satisfying the query. In this paper we only address the latter, but our solutions easily generalize to solve the MAP inference problem as well.

**Tuple-Independent Databases Revisited** Consider two possible tuples: $R(a_1), R(a_2)$, and the MLN consisting of features: $(R(a_1), w_1), (R(a_2), w_2)$. There are four possible worlds, $\emptyset, \{R(a_1)\}, \{R(a_2)\}, \{R(a_1), R(a_2)\}$, and their weights $\Phi(I_i)$ are:

$$1 \quad w_1 \quad w_2 \quad w_1 w_2$$

The partition function is $Z = 1 + w_1 + w_2 + w_1 w_2 = (1 + w_1)(1 + w_2)$. In this case the MLN defines a tuple-independent database, where the tuples $R(a_1), R(a_2)$ have probabilities

$p_1 = w_1/(1 + w_1)$, and $p_2 = w_2/(1 + w_2)$. Indeed, the four possible worlds have probabilities

$$(1 - p_1)(1 - p_2) \quad p_1(1 - p_2) \quad (1 - p_1)p_2 \quad p_1 p_2.$$

and this distribution is equivalent to the above (up to the multiplicative factor $Z$). More generally, we define:

DEFINITION 2. *A tuple-independent database, INDB, is a pair $\mathbf{D}_0 = (\mathbf{Tup}_0, \mathbf{w}_0)$ where $\mathbf{Tup}_0$ is a set of possible tuples and $\mathbf{w}_0(t)$ associates a real number to each tuple $t$.*

This definition is equivalent to the one given earlier, by setting the tuple probability to $p(t) = \mathbf{w}_0(t)/(1 + \mathbf{w}_0(t))$. Note that in a tuple-independent database a weight, $w$, represents the odds, $w = p/(1 - p)$. In other words, weight values of $0, 1, \infty$ correspond to probabilities $0, 1/2, 1$ respectively. From now on, unless otherwise stated, we will assume in the rest of this paper that an INDB is given as in Def. 2, that is, we are given the weights of the tuples, not their probabilities. The tuple's probability can always be recovered as $w/(1 + w)$.

## 2.4 MVDBs

In this section we introduce Markov Views (MARKOVIEW) and Markov View Databases (MVDB).

DEFINITION 3. *A MARKOVIEW is a rule of the form:*

$$V(\bar{x})[w_{expr}] \text{:-} Q \quad (3)$$

*where $V$ is the view name, $Q$ is a UCQ, $\bar{x}$ are head variables, and $w_{expr}$ is an expression representing a non-negative weight.*

Let $\mathbf{R}$ be a relational schema. *An* MVDB *is a triple* $(\mathbf{Tup}, \mathbf{w}, \mathbf{V})$, *where $\mathbf{Tup}$ is a set of possible tuples over the schema $\mathbf{R}$, $\mathbf{w} : \mathbf{Tup} \to [0, \infty]$ associates a weight to each possible tuple, and $\mathbf{V}$ is a set of* MARKOVIEWs.

Let $I_{\text{poss}}$ denote the deterministic database instance over the schema $\mathbf{R}$ consisting of all possible tuples $\mathbf{Tup}$ (forgetting their probabilities). For each MARKOVIEW $V$, denote $\mathbf{Tup}_V$ the result of evaluating $V$ on $I_{\text{poss}}$, and let $\mathbf{Tup}_\mathbf{V} = \bigcup_V \mathbf{Tup}_V$: this is the set of all possible tuples in all views. For each $t \in \mathbf{Tup}_V$, let $\mathbf{w}_V(t)$ denote its weight, as computed according by the view $V$.

The *semantics* of an MVDB is a restricted MLN having one feature $(F_t, w_t)$ for each tuple $t \in \mathbf{Tup} \cup \mathbf{Tup}_V$, defined as follows. For each possible tuple $t \in \mathbf{Tup}$ in the probabilistic database we associate the feature where the formula is $F_t = t$, the grounded atom represented by the tuple $t$, and the weight is $w_t = \mathbf{w}(t)$. For each possible tuple $t \in \mathbf{Tup}_V$ in a view $V$, consider the view definition $V(\bar{x})[w_{\text{expr}}]\text{:-}Q$: we associate $t$ with the feature where the formula is $F_t = Q(t)$, the Boolean query obtained by substituting the head variables $\bar{x}$ with the tuple $t$, and the weight is $\mathbf{w}_V(t)$. Thus, the feature $F_t$ is a ground tuple in the first case, and a Boolean UCQ in the second case. In both cases, $F_t$ has no free variables, and therefore it is already "grounded" according to the terminology of MLN's.

DEFINITION 4. *Let $(\mathbf{Tup}, \mathbf{w}, \mathbf{V})$ be an MVDB. Its semantics is given by the probabilistic database $\mathbf{D}_L$ (Def. 1) associated to the MLN $L = \{(F_t, w_t) \mid t \in \mathbf{Tup} \cup \mathbf{Tup}_\mathbf{V}\}$.*

We denote $P(Q)$ the probability of a Boolean query $Q$ on the probabilistic database associated to the MVDB. The problem in this paper is to compute $P(Q)$.

## 2.5 Discussion and Examples

An MVDB generalizes tuple-independent databases. Any INDB, $(\mathbf{Tup}, \mathbf{w})$ is in particular a MVDB, $(\mathbf{Tup}, \mathbf{w}, \emptyset)$, without any MARKOVIEWs. But MVDBs are much more powerful than INDBs, because they can impose correlations between arbitrary sets of tuples. We illustrate with several examples.

EXAMPLE 1. *Consider the* MVDB *with two possible tuples*, $\mathbf{Tup} = \{R(a), S(a)\}$, *with weights $w_1, w_2$ respectively, and a single* MARKOVIEW:

$$V(x)[w] \text{:-} R(x), S(x)$$

*Here $w$ is a constant. Intuitively, the view asserts that the tuples in $R$ and $S$ are correlated, by some weight $w$. We show now the associated MLN, $L$. There is a single tuple in the* MARKOVIEW, $\mathbf{Tup}_V = \{V(a)\}$, *and therefore the MLN has three features:*

$$L = \{(R(a), w_1), (S(a), w_2), (R(a) \wedge S(a), w)\}$$

*The probabilistic database $\mathbf{D}_L$ has four possible worlds,* $\emptyset, \{R(a)\}, \{S(a)\}, \{R(a), S(a)\}$, *with weights:*

$$1 \quad w_1 \quad w_2 \quad ww_1w_2$$

*Therefore, the two tuples $R(a), S(a)$ are correlated. When $w = 0$, then $R(a)$ and $S(a)$ are exclusive events; when $w = 1$ then they are independent events; when $w = \infty$ then both are certain tuples. More generally, when $w < 1$ then $R(a), S(a)$ are negatively correlated, and when $w > 1$ they are positively correlated.*

EXAMPLE 2. *A more complex example is $V(x)[w] = R(x), S(x, y)$. Each tuple $t = V(a)$ in the view defines the MLN feature $F_t = \exists y. R(a), S(a, y)$. The lineage of this Boolean query is $(R(a) \wedge S(a, b_1)) \vee (R(a) \wedge S(a, b_2)) \vee \ldots$, and the* MARKOVIEW *introduces a correlation between all tuples in the lineage expression. Here the* MARKOVIEW *introduces a correlation between a large number of tuples, in turn forming a large clique in the associated Markov Network. The view V1 in Fig. 1 is of this type, because the year attribute of $\texttt{Student}^p$ is projected out.*

EXAMPLE 3. *An example of a large* MVDB *is given in Fig. 1. The set of possible tuples $\mathbf{Tup}$ are defined by the deterministic tables at the top, and by the three queries in the middle of Fig. 1 (which also define the weight function $\mathbf{w}$ for the probabilistic tables). The* MARKOVIEWs *are V1, V2, V3 at the bottom of Fig. 1. The* MVDB *has over 6M probabilistic tuples (correlated) and over 6M deterministic tuples.*

Thus, MARKOVIEWs allow us to express both positive and negative correlations between probabilistic tuples. They are, however, strictly less expressive than MLNs, because they only allow UCQ as features. The advantage of imposing such a restriction is that it allows us to translate query evaluation of UCQs on MVDBs to query evaluation of UCQs on tuple-independent databases, as we show in the next section. For example, MARKOVIEWs cannot express a feature like "transitively closed", which would be written in MLN's like: $((R(x, y), R(y, z) \Rightarrow R(x, z)), w)$. One can express this in MARKOVIEWs if we extend them to allow negations:

```
V(x,y,z)[1.0/w] :- R(x,y),R(y,z),not R(x,z)
```

While the MLN rewards every grounding of $R(x, y), R(y, z) \Rightarrow R(x, z)$ by a factor $w$, the MARKOVIEW penalizes every violation by a factor $1/w$: the two features are equivalent.

The technique that we describe in the next section applies to this view too, i.e. queries can still be translated to tuple-independent database; the problem is that the new query contains negation, which is less well understood in probabilistic databases. For that reason, we are currently restricting MARKOVIEWS to be UCQs, without negation.

## 3. TRANSLATING MVDB TO INDB

### 3.1 Main Idea

Consider two possible tuples $R(a), S(a)$ with weights $w_1, w_2$ and the MARKOVIEW $V(x)[w] :- R(x), S(x)$, where $w$ is a constant. The four possible worlds have weights:

$$1 \quad w_1 \quad w_2 \quad ww_1w_2$$

We show how to reduce the probability $P(Q)$ to the probability of a query on a tuple-independent databases. Any Boolean query $Q$ corresponds to a subset of worlds; there are $2^4 = 16$ inequivalent Boolean queries. Then $P(Q) = \Phi(Q)/Z$, where $\Phi(Q)$ is the sum of the weights corresponding to the worlds that satisfy $Q$. For a very concrete example, if $Q = R(a) \vee S(a)$ then $\Phi(Q) = w_1 + w_2 + ww_1w_2$, and $P(Q) = (w_1 + w_2 + ww_1w_2)/(1 + w_1 + w_2 + ww_1w_2)$.

Consider now a tuple-independent database over three relations, $R, S, NV$, with three possible tuples $R(a), S(a), NV(a)$; their weights are $w_1, w_2, w_0$, where $w_1, w_2$ are as above and $w_0$ will be determined shortly. Consider the hard constraint $\neg W$, where $W \equiv R(a) \wedge S(a) \wedge NV(a)$. If one defines $V(a) = \neg NV(a)$, then $\neg W \equiv (R(a), S(a) \Rightarrow V(a))$. Seven out of the eight possible worlds satisfy $\neg W$, and their weights are:

| $\neg NV(a)$ | 1 | $w_1$ | $w_2$ | $w_1w_2$ |
|---|---|---|---|---|
| $NV(a)$ | $w_0$ | $w_0w_1$ | $w_0w_2$ | - |
| Total: | $1+w_0$ | $(1+w_0)w_1$ | $(1+w_0)w_2$ | $w_1w_2$ |

We write $\Phi_0$, $Z_0$ ($= \Phi_0(\texttt{true})$) and $P_0$ for the weight function, the partition function, and the probability defined by this tuple-independent database. Suppose we want to compute $P(Q)$ for some query $Q$ over the schema $R, S$, and consider its weight $\Phi_0(Q \wedge \neg W)$ in the new database: each column in the table above is either entirely included in the sum or is not included at all, because $Q$ does not refer to $NV$, only to $R$ and $S$. Thus, $\Phi_0(Q \wedge \neg W)$ is a sum of a subset of the weights in the last row, labeled "Total". Set $w_0$ such that $w = 1/(1+w_0)$: then[4] $\Phi_0(Q \wedge \neg W) = (1+w_0) \cdot \Phi(Q)$. For example, if $Q = R(a) \vee S(a)$ then:

$$\Phi_0(Q \wedge \neg W) = (1+w_0)w_1 + (1+w_0)w_2 + w_1w_2$$
$$= (1+w_0) \cdot (w_1 + w_2 + \frac{1}{1+w_0}w_1w_2)$$
$$= (1+w_0) \cdot \Phi(Q)$$

Therefore:

$$P(Q) = \frac{\Phi(Q)}{\Phi(\texttt{true})} = \frac{\Phi_0(Q \wedge \neg W)}{\Phi_0(\neg W)}$$
$$= \frac{P_0(Q \wedge \neg W)}{P_0(\neg W)} = \frac{P_0(Q \vee W) - P_0(W)}{1 - P_0(W)}$$

---

[4]$\Phi(Q)$ is a sum of a subset of weights in $1, w_1, w_2, ww_1w_2$ while $\Phi_0(Q \wedge \neg W)$ is a sum of the corresponding subset of weights in $(1+w_0), (1+w_0)w_1, (1+w_0)w_2, w_1w_2,$.

In summary, we have reduced the problem of evaluating $P(Q)$ on an MVDB to the problem of evaluating the probabilities $P_0(Q \vee W)$ and $P_0(W)$ on a tuple-independent database. Note that we can also express it as a conditional probability, $P(Q) = P_0(Q|\neg W)$; we prefer to use the expression above because both probability expressions $P_0(Q \vee W)$ and $P_0(W)$ are for UCQ's, for which query evaluation on tuple-independent databases is very well understood.

### 3.2 Main Theorem

DEFINITION 5. *Consider an MVDB $\mathbf{D} = (\mathbf{Tup}, \mathbf{w}, \mathbf{V})$ over the relational schema $\mathbf{R}$. Let $\mathbf{Tup_V}$ be the set of all possible tuples in all views, and $\mathbf{w}_V : \mathbf{Tup_V} \to [0, \infty]$ be their weights (Sect. 2.4).*

*Let $\mathbf{NV}$ denote the relational schema having one relation symbol $NV_i$ for each MARKOVIEW $V_i$.*

*The tuple-independent database associated to $\mathbf{D}$ is the following database over the schema $\mathbf{R} \cup \mathbf{NV}$: $\mathbf{D}_0 = (\mathbf{Tup}_0, \mathbf{w}_0)$, where the set of possible tuples and the weight function are defined by:*

$$\mathbf{Tup}_0 = \mathbf{Tup} \cup \mathbf{Tup}_{NV}$$
$$\mathbf{Tup}_{NV} = \{NV_i(\bar{a}) \mid V_i(\bar{a}) \in \mathbf{Tup}_{V_i}\}$$
$$\mathbf{w}_0(t) = \begin{cases} \mathbf{w}(t) & \text{if } t \in \mathbf{Tup}, \\ \frac{1-\mathbf{w}_V(t)}{\mathbf{w}_V(t)} & \text{if } t \in \mathbf{Tup_V} \end{cases}$$

In other words, to compute the INDB from the MVDB one proceeds as follows. All deterministic or probabilistic tables in MVDB become independent tables in the INDB, with the same weights. (A deterministic table in the MVDB has all weights $= \infty$, hence it remains a deterministic table in the INDB.) In addition, create a new relation $NV_i$ for each MARKOVIEW $V_i$: the possible tuples are all the possible tuples in the view $V_i$, and their weights are $w_0 = (1-w)/w$, where $w$ is the weight defined by the MARKOVIEW for that tuple. (Note that $w = 1/(1+w_0)$, a fact we will use in the proof.) The next theorem is the main theoretical result in this paper, and key to our technique. It says that, in order to compute UCQ queries on the MVDB, it suffices to compute UCQ queries over the associated INDB:

THEOREM 1. *Let $\mathbf{V} = \{V_1, \ldots, V_m\}$ be the MARKOVIEW in the MVDB. For $i = 1, m$ let $Q_i$ be the UCQ defining the view $V_i$. Denote $W_i$ the Boolean query*

$$W_i = \exists \bar{x}_i . NV_i(\bar{x}_i) \wedge Q_i(\bar{x}_i) \tag{4}$$

*Further define $W = \bigvee_i W_i$ (this is also a Boolean UCQ query). Then, for every Boolean query $Q$, the following holds:*

$$P(Q) = \frac{P_0(Q \vee W) - P_0(W)}{1 - P_0(W)} \tag{5}$$

PROOF. We follow the same steps as in Sect. 3.1. We start by computing $\Phi(Q)$ according to Def. 4:

$$\Phi(Q) = \sum_{J \subseteq \mathbf{Tup}} \Phi(J) = \sum_{J \subseteq \mathbf{Tup}} \Phi_0(J) \cdot \prod_{t \in \mathbf{Tup_V}: J \models F_t} \mathbf{w}(t)$$

Recall that the MLN associated to the MVDB has two sets of features $F_t$: for $t \in \mathbf{Tup}$ and for $t \in \mathbf{Tup_V}$. Thus, $\Phi(J)$ has two parts: $\prod_{t \in J} \mathbf{w}(t)$, which is the same as, $\Phi_0(J)$, the weight of $J$ in the INDB; and the product of all the weights of features satisfied by the world $J$. This justifies $\Phi(Q)$.

Next, we compute $\Phi_0(Q \wedge \neg W)$:

$$\Phi_0(Q \wedge \neg W) = \sum_{I : I \models \neg W} \Phi_0(I) \quad (6)$$

The possible world $I$ ranges over all subsets of $\mathbf{Tup} \cup \mathbf{Tup}_{NV}$, and, hence, it can be written as $I = J \cup K$, where $J \subseteq \mathbf{Tup}$ and $K \subseteq \mathbf{Tup}_{NV}$. Fix the $J$ component of $I$. Fix a MARKOVIEW $(V_i, w, Q_i(\bar{x}_i))$ and a possible tuple in this MARKOVIEW, $t = V_i(\bar{a}) \in \mathbf{Tup}_{V_i}$. There are two cases. Case 1: $\bar{a} \notin Q_i(J)$; equivalently, $J \not\models F_t$. In that case, we can satisfy the constraint $Q_i(\bar{x}_i) \Rightarrow \neg NV_i(\bar{x}_i)$, by either including or not including the tuple $t = NV_i(\bar{a})$ in $K$: the sum of the two weights is $1 + \mathbf{w}_0(t)$. Case 2: $\bar{a} \notin Q_i(J)$; equivalently $J \models F_t$. In that case, in order to satisfy the constraint we must remove the tuple $t$ in $K$, hence its multiplicative contribution is 1. Thus, we rewrite Eq. 6 grouping by $J$:

$$\sum_{J,K:J\cup K \models \neg W} \Phi_0(J \cup K) = \sum_J \Phi_0(J) \prod_{t:J \not\models F_t} (1 + \mathbf{w}_0(t))$$
$$\cdot \prod_{t:J\models F_t} 1 = P \cdot \sum_J \Phi_0(J) \cdot \prod_{t:J\models F_t} \frac{1}{1+\mathbf{w}_0(t)} = P \cdot \Phi(Q)$$

where $P = \prod_t (1 + \mathbf{w}_0(t))$. In the last line we used the fact that $1/(1 + \mathbf{w}_0(t)) = \mathbf{w}(t)$. The theorem follows now by noting that $P(Q) = \Phi(Q)/\Phi(\texttt{true})$, and repeating the argument at the end of Sect. 3.1. □

We end this section with three observations. First, we note that the lineage of $Q \vee W$ is no larger than the lineages of $Q$ and $W$ combined. In fact, the lineage of $Q \vee W$ is precisely the disjunction of the two lineage expressions of $Q$ and $W$ respectively. Thus, our translation does not add any complexity to the probabilistic inference problem. Second, we note that the translation gives us an immediate tool for identifying tractable cases. The UCQ queries that can be evaluated in PTIME over INDB are fully characterized in [8], and are called *safe* queries. An immediate corollary of Theorem 1 is that query evaluation over MVDB is tractable if both $Q \vee W$ and $W$ are safe. Other tractability results on INDB also carry over immediately to MVDBs, for example the query compilation results in [15]. Finally, a comment on *denial views*, which are MARKOVIEWS where the weight is 0: for example view V2 in Fig. 1 is a denial view. In that case NV is a deterministic table, since its weight is $(1 - 0)/0 = \infty$, and can be dropped entirely from the definition of $W_i$, simplifying Eq. 4.

### 3.3 Discussion: Negative Probabilities

Some of the probabilities in $\mathbf{D}_0$ may be negative: if $w > 1$, then $w_0 = (1-w)/w < 0$, and the probability $p_0 = w_0/(1+w_0) = 1 - w$ is negative. This may raise questions about the soundness of our approach. However, negative probabilities have already been considered before; it has been proven that probability theory can be consistently extended to allow for negative probabilities [2], and there is interest in applying them to quantum mechanics [6] and financial modeling [13]. In our setting, the negative probabilities have a much more benign role: they are simply numbers that need to be plugged into the right hand side of Eq. 5 to make the equality hold. Every query answer $P(Q)$ will be a correct probability, in $[0, 1]$, even if the probabilities $P_0$ on the right are negative. All familiar equalities still hold for $P_0$: for example, the rules for negation $P_0(\neg Q) = 1 - P_0(Q)$, and inclusion/exclusion $P_0(Q_1 \vee Q_2) = P_0(Q_1) + P_0(Q_2) - P_0(Q_1 \wedge Q_2)$ still hold; similarly, if $Q_1, Q_2$ are independent queries (their lineages have no common variables) then the laws of independence continue to hold: $P_0(Q_1 \wedge Q_2) = P_0(Q_1)P(Q_2)$ and $P_0(Q_1 \vee Q_2) = 1 - (1 - P_0(Q_1))(1 - P_2(Q_2))$; similarly, Shannon's expansion formula holds. In fact all exact inference methods (Davis-Putnam procedure, tree-width based methods, OBDD constructions) work without any modification on probability spaces with negative probabilities. We take advantage of this fact in the next section.

Approximate methods, however, no longer work out-of-the-box. For example, the inequality $P_0(Q_1 \vee Q_2) \leq P_0(Q_1) + P_0(Q_2)$ must be replaced with the weaker $|P_0(Q_1 \vee Q_2)| \leq |P_0(Q_1)| + |P_0(Q_2)|$; another issue is that, approximation methods may no longer return final values in $[0, 1]$; it is also unclear how to run sampling-based methods. In this paper we do not consider any approximation or sampling methods, instead restrict the discussion only to exact probability computation. We show next how to do this quite effectively.

## 4. COMPILING MARKOVIEWS

We have translated the problem $P(Q)$ on an MVDB into the problem $P_0(Q \vee W)$ on a tuple-independent database. Thus, from now on we will only consider tuple-independent databases.

Even though query evaluation over INDB has been well-studied, there is an important distinction in our setting. While the lineage of $Q$ is typically small, the lineage of $W$ is usually large, especially as we try to capture more correlations. For example, the DBLP data in Fig. 1 has over 6M probabilsitic tuples, and many of them (if not all) will be included in the lineage of $W$. Since $W$ is defined offline, we employ the strategy of pre-compiling $W$, in order to speed up query evaluation of $P_0(Q \vee W)$ online. In this section, we will discuss the data structure *MV-Index* into which we compile $W$ along with the algorithm for compilation. MV-Index has been designed to speed-up the online evaluation of $P_0(Q \vee W)$, which we discuss next.

Throughout this section we assume a tuple-independent probabilistic database $\mathsf{D}_0$. We associate to each tuple a Boolean variable $X_1, X_2, \ldots, X_n$, Sect. 2.1. We consider only Boolean queries in this section. $W$ is already a Booelan query (Eq. 4); the user query $Q$ is typically not a Boolean query, but for the purpose of query evaluation we substitute its head variables with an answer tuple, thus transforming it into a Boolean query. We denote $\Phi_Q$ the *lineage* of the query $Q$ on the probabilistic database. $\Phi_Q$ is a Boolean formula using variables $X_1, \ldots, X_n$ (see, e.g. [33]). Figure 3 shows a probabilistic database, a query $Q$, and its lineage expression $\Phi_Q$. Notice that the lineage of a disjunction is the disjunction of the lineages, $\Phi_{Q \vee W} = \Phi_Q \vee \Phi_W$.

### 4.1 MV Index

An MV-Index consists of a set of OBDD augmented with certain pre-computations and indices that we describe below. But first we briefly review OBDD.

**OBDD**: An Ordered Binary Decision Diagrams, OBDD [5] is a rooted DAG, where internal nodes are labeled with Boolean variables and have two outgoing edges, labeled 0 and 1; sink nodes (leaves) are labeled 0 or 1. There are two constraints: every path from the root to a leaf must visit each variable at most once, and any two paths must visit the variables in the same order (missing variables are OK), see an example in Figure 3. Given an OBDD for $\Phi$ one can compute the probability $P_0(\Phi)$ in linear time. Denote $p(u)$ the probability of the Boolean formula encoded by the OBDD rooted at node $u$. If $u$ is a leaf node, set $p(u) = 0$ or $p(u) = 1$ (depending on its label); otherwise it is labeled with some variable, say $X_i$, and has two children, say $u_0, u_1$ corresponding to the outgoing edges labeled with 0,1 respectively. We set $p(u) = (1 - P_0(X_i)) \cdot p(u_0) + P_0(X_i) \cdot p(u_1)$. This formula (Shannon expansion) also holds when $P_0(X_i)$ is negative. The size of an OBDD is the total number of nodes in it. The width at level $i$ is the number of nodes labeled with variable $X_i$ and width is the maximum width at any level. Note that the size of OBDD is at most the width times the number of variables.

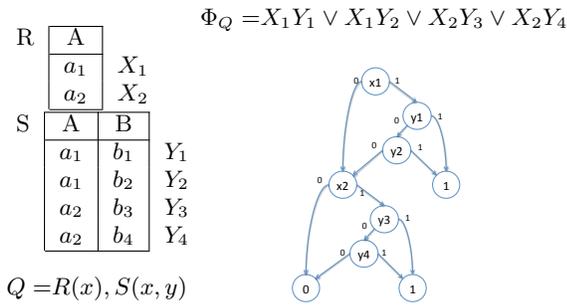

**Figure 3: A query $Q$ on a probabilistic database, its lineage $\Phi_Q$, and an OBDD for $\Phi_Q$**

**The MV-Index** An *augmented OBDD* for $\Phi$ is an OBDD where each node $u$ is annotated with two quantities. First, $u$.probUnder stores $p(u)$. Second, $u$.reachability stores the sum of the probability of all paths starting from root to $u$. The probability of a path is defined as a product of the following factors: for an edge $X_i = 1$ we include the factor $P_0(X_i)$, and similarly for $X_i = 0$ we include $(1 - P_0(X_i))$. To see the intuition, suppose we want to compute $P_0(\varphi \wedge \Phi)$ for some "small" expression $\varphi$, and assume $\varphi = X_1$, the root variable of the OBDD. Then we just return $p * u$.probUnder, where $p$ is the probability of $X_1$ and $u$ is its 1-child. Assume now $\varphi = X_i$, and that there are $c$ nodes in the OBDD labeled $X_i$, $u_1, u_2, \ldots, u_c$. Assume every path from the root to a sink contains one of these nodes. Let $v_1, v_2, \ldots, v_c$ be their 1-children. Then $P_0(X_i \wedge \Phi) = p * \sum_{j=1}^{c} u_j$.reachability $* v_j$.probUnder. When the width $c$ is small, then the augmented OBDD allows us to compute these probabilities efficiently.

Finally, an MV-Index consists of a set of augmented OBDD, each of them associated with a particular key. These OBDD are over disjoint set of variables. Besides the OBDD, it keeps two indices. *InterBddIndex*, given a tuple, returns the key of the OBDD where the tuple is located. *IntraBddIndex* returns all the nodes in the OBDD which correspond to that tuple. For e.g., in our previous example where we were computing $P(X_1)$, we still needed to locate all the nodes labeled with $X_1$. These indices enable us to do that in constant time.

## 4.2 Constructing the MV-Index

In this section, we will describe our algorithm to construct an OBDD for a UCQ. Given an OBDD, adding the precomputations and indices needed to make it an MV-Index are straightforward and hence skipped in this section.

Let $\Pi$ be the order in which an OBDD visits the variables on each path. (E.g. $\Pi = X_1, Y_1, Y_2, X_2, Y_3, Y_4$ in Fig. 3). The order $\Pi$ uniquely determines the OBDD [35], up to merging of equivalent nodes, therefore the problem of finding a small OBDD is equivalent to finding a good order $\Pi$ (meaning one that leads to a small OBDD). Denote $OBDD_\Pi(\Phi)$ the OBDD of $\Phi$ with order $\Pi$. If $\Phi = \Phi_1 \vee \Phi_2$, or $\Phi = \Phi_1 \wedge \Phi_2$, and we are given OBDDs $G_1, G_2$ for $\Phi_1, \Phi_2$ with the same order $\Pi$, then one can compute $OBDD_\Pi(\Phi)$ in time $O(|G_1||G_2|)$ by a procedure called *synthesis*. CUDD [32], a widely popular package for OBDDs, uses this synthesis procedure. It starts with some order $\Pi$ and synthesizes the OBDD traversing $\Phi$ recursively.

We do the following improvement over the synthesis procedure in CUDD. Suppose $\Phi_1, \Phi_2$ are independent, i.e. they use disjoint sets of Boolean variables, then one can synthesize $\Phi_1 \vee \Phi_2$ more efficiently: stack the OBDD's $G_1, G_2$ on top of each other, and redirect every 0-labeled leaf of $G_1$ to the root of $G_2$ (for $\Phi_1 \wedge \Phi_1$ one would redirect the 1-labeled leaves). We call this *concatenation*. The size of the new OBDD is only $|G_1| + |G_2|$, and, unlike synthesis, concatenation is a constant time operation. Thus, concatenation represents a major improvement over synthesis. We explain next where we can use concatenation when computing the OBDD of a UCQ.

Consider a Conjunctive Query(CQ) $Q$. A *root variable* in $Q$ is a variable $x$ that appears in all atoms of $Q$. One can write $Q \equiv Q[a_1/x] \vee Q[a_2/x] \vee \ldots$, where $a_1, a_2, \ldots$ form the active domain of $x$. It is not hard to see that if $Q$ has no self-joins then $Q[a_i/x]$ and $Q[a_j/x]$ have no tuples in common; hence the OBDD of $Q$ can be obtained by concatenating the OBDD of $Q[a_i/x]$. This has been illustrated for $Q = R(x), S(x, y)$ in Fig. 3.

In general for a UCQ $Q = Q_1 \vee Q_2 \vee \ldots$, where $Q_i$ are CQ, let $x_i$ be a root variables of $Q_i$. We write $Q = \exists x_1.Q_1 \vee \exists x_2.Q_2 \vee \ldots = \exists z.(Q_1[z/x_1] \vee Q_2[z/x_2] \vee \ldots)$. The new variable $z$ occurs in all atoms of all conjunctive queries. We call $z$ a *separator variable* if any two atoms with the same symbol contain $z$ on the same attribute position. Let $a_1, a_2, \ldots$ be the active domain for $z$. Then once again the same property holds: $Q \equiv Q[a_1/z] \vee Q[a_2/z] \vee \ldots$, and the queries on the right are independent, i.e. they do not have any common tuples. For e.g., let $Q = R(x_1), S(x_1, y_1) \vee T(x_2), S(x_2, y_2)$. Both $x_1$ and $x_2$ are root variables, and we write the query as follows (showing quantifiers explicitly now):

$$Q = \exists z.[\exists y_1.(R(z) \wedge S(z, y_1)) \vee \exists y_2.(T(z) \wedge S(z, y_2))]$$

Thus, $Q = Q[a_1/z] \vee Q[a_2/z] \vee \ldots$ and the OBDD for $Q$ has a size that is the sum of the OBDD for $Q[a_i/z]$. In general

PROPOSITION 1. *If $z$ is a separator in $Q$, then there exists an OBDD for $Q$ of size at most the sum of OBDD of $Q[a_i/z], i = 1, n$, where $a_1, a_2, \ldots, a_n$ is the active domain.*

Finally, consider $Q = R(x_1), S(x_1, y_1) \vee S(x_2, y_2), T(y_2)$: this query does not have a separator variable. For such queries we fall back on general tools like CUDD. Then there are hybrid cases like $Q = Q_1 \vee Q_2$, $Q_1 = R(x), S(x, y), R(y)$, $Q_2 = S(x, y)T(x)$. Here only $Q_2$ has a root variable, so we will choose $\Pi$ such that we can compute $OBDD_\Pi(Q_2)$ by concatenation, then use synthesis to compute $OBDD_\Pi(Q_1)$ using the same order $\Pi$. Having reviewed the concepts from our prior work, we are now ready to describe the construction, which is the novel part of this section.

Fix a relational schema $\mathbf{R} = \{R_1, \ldots, R_k\}$. Order the relation names from smaller to larger arities. Let $\pi = \{\pi_{R_1}, \ldots, \pi_{R_k}\}$ be set where each $\pi_i$ is a permutation on the attributes of the relation $R_i$. That is, if $arity(R_i) = m$ then $\pi_i$ is any permutation on $1, \ldots, m$. Consider a database instance $I$, over an ordered active domain $a_1 < a_2 < \ldots < a_n$. Then $\pi$ defines an order $\Pi$ on all tuples in $I$, as follows. $\Pi = \Pi_1, \Pi_2, \ldots, \Pi_n$, where each $\Pi_j$ is the order obtained recursively as follows: for each relation $R_i$, retain from $D$ only those tuples where the first attribute in $R_i$ (according to $\pi_{R_i}$) is $a_j$, then project out that attribute, and compute $\Pi_j$ recursively on the smaller database. For example, consider the schema $R(A), S(A, B)$, and the permutation $\pi_R = (A)$ and $\pi_S = (A, B)$. Consider the database instance in Fig. 3), where the active domain is ordered by $a_1 < a_2 < b_1 < b_2 < b_3 < b_4$. Then $\Pi = X_1, Y_1, Y_2, X_2, Y_3, Y_4$.

Let $\pi$ be given and let $Q$ be a query. If $x, y$ are two variables then we write $x >_\pi y$ if whenever $y$ occurs in an atom on some attribute $B$, then $x$ also occurs in that atom on some attribute $A$, and $A$ comes before $B$ in the permutation $\pi_{R_i}$.

Given $\pi$ and a query $Q$, the following recursive procedure $ConOBDD(\pi, Q)$ constructs $OBDD_\Pi(Q)$, where $\Pi$ is the permutation on the tuples associated to $\pi$:

- **R1** $Q = Q_1 \vee Q_2$ : if $Q_1, Q_2$ have no relations in common, concatenate, else synthesize.

- **R2** $Q = Q_1 \wedge Q_2$ : if $Q_1, Q_2$ have no relations in common, concatenate, else synthesize.

- **R3** $Q = \exists x.Q_1$ : if $x >_\pi y$ for all variables $y$ in $Q_1$, then concatenate, else synthesize.

- **R4** $Q = R(\bar{a})$ : trivial

We choose heuristically $\pi$ such as to minimize the number of synthesis steps in **R3**. In particular, if $Q$ has a separator variable, then we always choose $\pi$ such that every attribute holding a separator variable occurs first in the permutation $\pi$. Call a query $Q$ *inversion-free* if there exists $\pi$ such that only concatenations are performed in **R3**. (This is equivalent to the definition of inversion-free queries in [15].) Let $\{a_1, \ldots, a_n\}$ be the active domain. The following proposition, based on [15], gives guarantees on the size of $ConOBDD(\pi, Q)$ in certain cases.

PROPOSITION 2. *Let $N$ be the size of the OBDD returned by $ConOBDD(\pi, Q)$. (a) If $Q$ admits a separator variable $z$, then $N = \sum_j N_j$, where $N_j$ is the size of the OBDD of $Q[a_j/z]$, $j = 1, n$. (b) If $Q$ is inversion-free, then the OBDD has constant width; hence $N = O(n)$. [15].*

### 4.3 Querying an MV-index

Our goal is to compute efficiently Eq.5, whose numerator is $P_0(Q \vee W) - P_0(W) = P_0(Q \wedge \neg W)$. The OBDD for $\neg W$ is obtained immediately from the OBDD for $W$, by switching the 0 and 1 sink nodes. In this section we will show how to use an MV-Index for $\neg W$, to compute efficiently $P_0(Q \wedge \neg W)$: we call this operation *intersection*. Denote $G_W = OBDD_\Pi(\neg W)$. Given $Q$, we first construct $G_Q = OBDD_\Pi(Q)$. Note that, although $\Pi$ is imposed by $W$, constructing $G_Q$ is usually quite efficient, because lineage of $Q$ is typically small. A naive next step would be to compute $G_Q \wedge G_W$, but this requires traversing the entire index. We briefly review our improvements over the naive algorithm next.

**MVIntersect** MVIntersect uses $G_Q = OBDD_\Pi(Q)$ to guide a search through $G_W = OBDD_\Pi(\neg W)$ and sum the probability of only those worlds that satisfy $Q$ via a top-down algorithm. One of the main challenges in doing a top-down intersection is maintaining a cache for memoization. This is well-studied(c.f. [14] for a top-down algorithm), so we don't discuss it here. Since $G_W, G_Q$ have same variable order, our algorithm just traverses $G_W$ and prunes out branches where $G_Q$ is false. To do this we could implement a DFS traversal of $G_W$ and in the stack, we maintain the nodes from both $G_W$ and $G_Q$. Whenever we pop false from $G_Q$, we don't traverse the subtree below. The stack here though can become very big and the node from $G_Q$ is almost always the same, since $G_Q$ is very small. MVIntersect exploits this by not adding the same node from $G_Q$ consecutively, but keeping a count of how many times the same query node has been pushed.

**CC-MVIntersect** Since our algorithm is main-memory and considering the increasing gap in the access time of cache and memory, we optimized our data structure to be more cache-conscious and minimize random accesses. A typical BDD data structure, for instance used in CUDD, is to store bdd nodes as pointers and each node contains the pointers to its neighbors. We improve upon it by keeping the bdd nodes in a vector sorted according to the DFS traversal of the obdd. Querying the obdd now can be done by a sequential traversal of this vector. The new bdd nodes though may need to store some additional information about their neighbors now. We call this approach CC-MVIntersect.

PROPOSITION 3. *Let $G_W$ be the OBDD for $\neg W$. Let $X_i$ and $X_j$ be the first and last Boolean variable (according to the permutation $\Pi$) occurring in the lineage of $Q$, and let $m = j - i + 1$. The CC-MVIntersect procedure runs in time $O(m \cdot w)$, where $w$ is the width of $G_W$.*

Finally, we would like to point out that if $W$ is inversion-free then since its OBDD is of constant width, the runtime of CC-MVIntersect is linear in the *span* of the query Q, i.e. the distance between the first and last variable in its lineage.

## 5. EXPERIMENTAL EVALUATION

We report here our experimental evaluation of MVDBs, on real data obtained from [19]. We addressed four questions. How do MARKOVIEWS, and indexed MARKOVIEWS compare to other approaches for probabilistic inference on large Markov Networks? How effective is the MV-index construction algorithm compared to the standard approach for constructing OBDDs? How effective is the MV-index-based query evaluation method, how significant is the improvement of the CC-MVIntersect algorithm over MVIntersect? And, finally, how do MVDBs scale to the entire DBLP dataset?

**Set Up** For probabilistic inference, we compare our approach with Alchemy [37], the de-facto standard inference engine for MLN. For OBDD construction, we have extended CUDD [32], a widey-used OBDD package; for the OBDD experiments we compare native CUDD with our obdd construction algorithm. Our implementation is written in C++.

We used Postgres 9.0 for our experiments which are run on a 2.66 GHZ Intel Core 2 Duo, with 4GB RAM, running Mac OS X 10.6. The dataset used is DBLP [19] and we consider features from Fig. 1.

## 5.1 Comparison with Alchemy

To compare against Alchemy, we construct an MLN using features over $\text{Advisor}^p$ and $\text{Student}^p$ i.e., $V_1, V_2$, Fig. 1. We did not use $\text{Affiliations}^p$, because the characters in the affiliation string violated Alchemy's requirement for constants. MLN do not allow features to have parameterized weights like in MVDBs, instead MLNs require a constant weight for each feature. One alternative is to materialize each feature into multiple features, one for each value of the weight; this would have increased the number of features. We opted instead to use constant weight in Alchemy, for simplicity. In MVDB we continued to use the weights exactly as described in Fig. 1.

Not surprisingly, Alchemy did not scale to the entire dblp dataset; we could scale it only up to $\text{aid} = 10,000$, in $\text{Student}^p(\text{aid,year})$, where aid ranges upto 1M. In our experiments, we generate 10 datasets, with domain of aid from 1 to $i*1000, i = 1\ldots 10$. The lineage size of the MARKOVIEW, i.e. tuples involved in the constraints (in other words the size of $\Phi_W$, Sect. 4) is plotted in Fig. 4. Over each such dataset we ran two queries of the form *find the advisor of some student X*, and *find all students of some advisor Y*. Fig. 5 and Fig. 6 show the running time comparison of Alchemy vs MARKOVIEWS. We show both the total execution time and also the time alchemy reports it spent in just sampling. It is known that Alchemy is very inefficient during the grounding phase, and that database techniques can speed up this phase considerably [20], therefore the interesting line in Fig. 5 is the lower line, *Alchemy-sampling*, which is Alchemy's reported sampling time, and is a better measure (likely a lower bound) on the total probabilistic inference time. The sampling method used is MC-Sat[25].

The results in Fig. 5 and Fig. 6 show that Alchemy's MC-Sat algorithm is comparable (within a factor of 5, up or down) with evaluating the MARKOVIEWS directly by constructing the OBDD. Note that OBDDs are an exact inference method while MC-Sat is an approximate method; and, also, OBDDs are not ideal for pure probabilistic inference, but are more appropriate for compilation. Once we use them for this purpose, by constructing an MV-index, the performance of the MARKOVIEWS increases dramatically, and remains mostly constant as we increase the data size.

## 5.2 Comparison with CUDD

Here we studied the OBDD construction time, and compared it to the OBDD construction time as performed by native CUDD. The running time depends on the size of the resulting OBDD, and therefore we needed a feature that allowed us both to increase the size of the OBDD linearly, and for which the OBDD constructed by native CUDD was the same as that resulting from our optimization. The MARKOVIEW V2 in Fig. 1 had both properties. As we varied linearly the domain of aid1 in $\text{Advisor}^p(\text{aid1,aid2})$ from [1000] to [10000], the size of the resulting OBDD varied linearly, as shown in Fig. 7. Furthermore, we checked that the size of the OBDDs returned by the two methods were indeed the same. Figure 8 shows that our algorithm was two orders of magnitude more efficient than CUDD's OBDD construction. Since aid1 is a separator, our approach exploits concatenation, which is much more efficient than standard OBDD synthesis used by CUDD. Since CUDD is effective at detecting equivalent nodes in the OBDD, it constructs the same OBDD, but it takes a lot of time to achieve the same result. We could not use CUDD to construct an OBDD for the entire DBLP dataset, even for V2: we estimate it would have taken several hours.

## 5.3 MV-index-based Query Evaluation

Recall that our query evaluation $P_0(Q \vee W)$ is based on optimized intersection of the OBDDs for $Q$ with the MV-index for $W$. Here we compare two algorithms: MVIntersect and CC-MVIntersect (recall that that CC stands for cache-conscious). We used the same setting as in previous section. We used a simple query $Q$ whose lineage consisted of 20 tuples chosen as a worst case scenario: it forced the system to traverse entire MV-index, rendering all pre-computations and indices useless. Thus, query evaluation requires a complete intersection of the two OBDDs. Figure 9 shows the running time of the two algorithms. As expected, the running time varies linearly in the size of the MV-index (which, recall Fig. 7, we designed carefully to be linear in the size of the data), because the entire OBDD must be traversed. The cache-conscious improves by a factor of 2 over the plain algorithm. We note that this is the worst case scenario; in typical cases, the query needs to traverse only a fraction of the MV-index, as will become clear next.

## 5.4 Scalability to a Large Dataset

We finally report our scalability results on the entire DBLP dataset, as described in Fig. 1. The MARKOVIEWS have a separator, hence the MV-index is obtained by concatenating many small OBDDs; their total size is 1.38M. Note that not all probabilistic tuples ended up in the MV-index, because some did not participate in any views. It took under one hour to construct the OBDD and index.

We evaluated 10 queries, of form *find all students of an advisor X*, and *find the affiliation of a person Y*. The running times are reported in Fig. 10, Fig. 11 respectively. In all cases we used the CC-MVIntersect. As one can see, the running times are below 5ms for all queries, and many are below 1ms. Query evaluation time includes the round trip call to Postgres, to compute the query's lineage, then the time to access the OBDD index, which is a main memory data structure. Since each query includes a constant (the name of the advisor X, or the name of the person Y), only a small portion of the full OBDD had to accessed at runtime, which explains the performance of query evaluation. Note that all probability computations are exact: this is unlike Alchemy's which are approximate.

## 5.5 Discussion

We have demonstrated how MV-indexes can be used to dramatically speed up query evaluation. The key ingredient that makes the index construction possible is the translation of the query evaluation from a Markov Network to a tuple-independent database, since OBDDs are only possible over independent variables. The OBDD construction is non-trivial. However, once the index is constructed, the performance of query evaluation becomes comparable to evaluating that query in postgres.

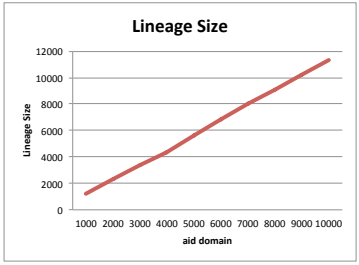
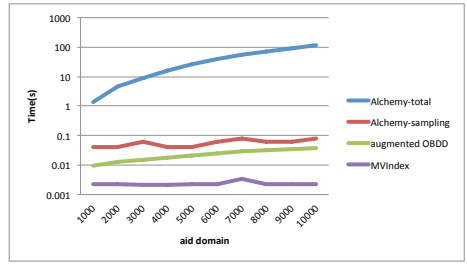
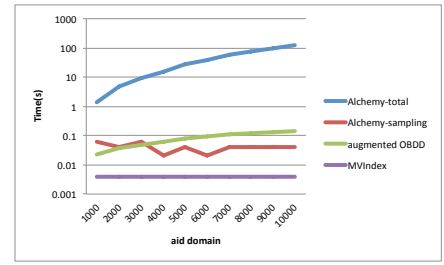

Figure 4: Lineage Size of MV for each dataset

Figure 5: Alchemy vs MV for querying advisor of a student

Figure 6: Alchemy vs MV for querying all students of an advisor

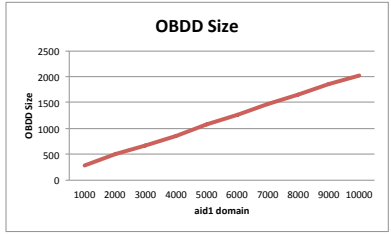
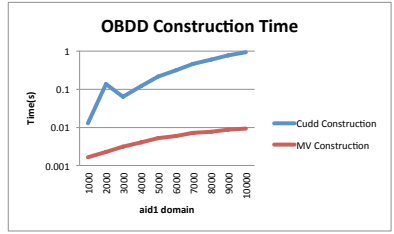
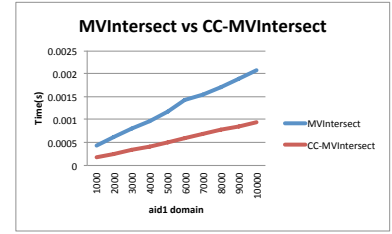

Figure 7: OBDD Size

Figure 8: Cudd vs MV : OBDD construction time

Figure 9: Querying time : MV-Intersect vs CC-MVIntersect

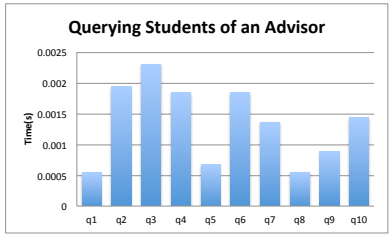
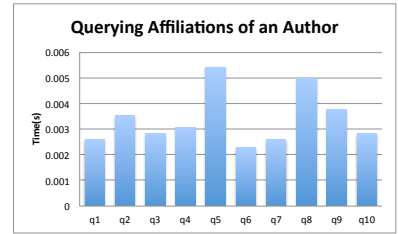

Figure 10: Querying students of an Advisor

Figure 11: Querying Affiliation

Future work is needed to assess the applicability of MVDBs, both in terms of modeling power and in terms of performance. Typical applications of MLN's [26, 31, 27] use 4-10 features, most of which can already be expressed as MARKOVIEWS, but they use much smaller data sets.

## 6. RELATED WORK

**Query Evaluation on Tuple Independent Databases.** This problem reduces to that of evaluating the probability $P(\Phi)$ of a Boolean formula $\Phi$ (over the Boolean variables $X_t$) called the *lineage* of the query $Q$. There are two lines of research on query evaluation on INDB's. One aims at identifying classes of queries for which $P(Q)$ can be computed in polynomial time in the size of the database: these are called *safe queries*. For UCQ's without inequality predicates (like $x < 5$ or $x \neq y$), there exists a syntactic characterization of safe queries that is complete, i.e. a dichotomy: either $Q$ is safe and then one can compute $P(Q)$ in PTIME, or $Q$ is unsafe and then computing $P(Q)$ is #P-hard. The other line of research aims at developing effective heuristics for computing $P(\Phi)$. In this formulation the problem is related to model counting, for which several effective heuristics exists. The most popular is the Davis-Putnam procedure, which is based on Shannon expansion, see for example [3], and which has been used in probabilistic databases in [17];

refinements of this procedure also exist [10]. Based on ideas similar to the Davis-Putnman procedure, a number of techniques have been described for using OBBDs for query evaluation on probabilistic databases [21, 22, 15]. Finally, these evaluation methods have been extended with several approximation techniques [24, 11].

**Markov and Bayesian Networks in Probabilistic Databases.** Several proposals exists for extending probabilistic databases to represent Markov Networks or Bayesian Networks. For example, in [16] the probabilistic database is defined directly as a Markov Network. The system uses a novel indexing techniques for the junction tree decomposition of the network, allowing queries on a large database to be evaluated efficiently at runtime. The method works very well, but only when the tree width of the Markov Network is small, otherwise the method is intractable. Note that the Markov Networks in MVDBs have very large cliques, hence very large tree widths. Tuffy [20] implements MLN's directly in a relational database system. Its focus is less on developing new algorithms, instead it is on leveraging query processing in the database in order to implement existing MLN algorithms. Two other projects [36, 34] implement MCMC inside a relational database for efficiently answering general SQL queries over a CRF model, commonly used in Information Extraction. The approach taken by these systems is to scale up existing general purpose probabilistic inference methods. Our MVDB approach differs from these in that we do not optimize any existing algorithm for inference over complex probabilistic models, but propose a new approach by which we translate from a complex probabilistic model to a tuple-independent probabilistic model.

## 7. CONCLUSION

We described a new approach to probabilistic databases, which allows complex correlations to be defined between the tuples in a database. Our new approach is based on MARK-OVIEWS, which are a restricted form of a Markov Logic Network feature. We made two contributions that allow queries to be processed very efficiently on such databases. The first is a translation from MARKOVIEWS into tuple-independent databases. The second is a compilation of the MARKOVIEWS into OBDDs, which dramatically speeds up query execution. We have also validated our techniques experimentally.